\documentclass[12pt]{iopart}
\usepackage{graphicx}
\usepackage{iopams} 
\usepackage{amstext,amssymb}
\usepackage{booktabs}
\usepackage{cite}

\newcommand{\yso}{Y$_2$SiO$_5$}
\newcommand{\Prs}{$\text{Pr}^{3+}\text{:Y}_2\text{SiO}_5$}
\newcommand{\Eurp}{$\text{Eu}^{3+}\text{:Y}_2\text{SiO}_5$}

\begin{document}

\title{Multimode capacity of atomic-frequency comb quantum memories}

\author{Antonio Ortu$^1$, Jelena V. Rakonjac$^2$, Adrian Holz\"{a}pfel$^1$, Alessandro Seri$^2$, Samuele Grandi$^2$, Margherita Mazzera$^{3}$, Hugues de Riedmatten$^{2,4}$, and Mikael Afzelius$^1$}

\address{$^1$Department of Applied Physics, University of Geneva, CH-1211 Geneva 4, Switzerland \\
$^2$ICFO - Institut de Ciencies Fotoniques, The Barcelona Institute of Science and Technology, Castelldefels (Barcelona) 08860, Spain \\
$^3$Institute of Photonics and Quantum Sciences, SUPA, Heriot-Watt University, Edinburgh EH14 4AS, UK \\
$^4$ICREA-Instituci\'o Catalana de Recerca i Estudis Avan\c cats, 08015 Barcelona, Spain}

\ead{samuele.grandi@icfo.eu}
\ead{mikael.afzelius@unige.ch}

\vspace{10pt}

\begin{abstract}

Ensemble-based quantum memories are key to developing multiplexed quantum repeaters, able to overcome the intrinsic rate limitation imposed by finite communication times over long distances. Rare-earth ion doped crystals are main candidates for highly multimode quantum memories, where time, frequency and spatial multiplexing can be exploited to store multiple modes. In this context the atomic frequency comb (AFC) quantum memory provides large temporal multimode capacity, which can readily be combined with multiplexing in frequency and space. In this article, we derive theoretical formulas for quantifying the temporal multimode capacity of AFC-based memories, for both optical memories with fixed storage time and spin-wave memories with longer storage times and on-demand read out. The temporal multimode capacity is expressed in key memory parameters, such as AFC bandwidth, fixed-delay storage time, memory efficiency, and control field Rabi frequency. Current experiments in europium- and praseodymium-doped \yso{} are analyzed within this theoretical framework, and prospects for higher temporal capacity in these materials are considered. In addition we consider the possibility of spectral and spatial multiplexing to further increase the mode capacity, with examples given for both rare earh ions. 

\end{abstract}


\section{Introduction}
\label{sec_intro}

The realization of quantum networks relies on the distribution of entanglement over remote quantum nodes using photons. In ground-based networks the photons travel between the nodes in optical fibers, which causes the entanglement rate to decrease exponentially with the distance. The loss is due to optical fiber attenuation, which limits ground-based and repeater-less entanglement distribution schemes to a few hundred km \cite{Inagaki2013}, while still allowing quantum key distribution using weak coherent states up to about 600 km \cite{Boaron2018a,Lucamarini2018,Minder2019,Chen2021,Pittaluga2021}. 

To overcome this limitation, quantum repeaters have been proposed \cite{Briegel1998,Duan2001,Simon2007,Sangouard2009,Sangouard2011}. Near-term quantum repeaters are based on creating heralded entanglement within elementary links. These are the individual segments in which the network branch is divided into. Each elementary link has two so-called ``quantum nodes'', with the ability to generate entanglement and store a part of it in a quantum memory, while the other part is used to perform entanglement swapping operations with neighboring links. Repeating this swapping operation through the whole chain of elementary links will eventually lead to entanglement between the two end nodes of the network branch. Heralding and storing the entanglement between remote quantum memories in an elementary link is the key to the sub-exponential scaling of entanglement distribution rate with distance in quantum repeaters. 

The generation of remote heralded entanglement usually relies on a measurement-induced process that requires the detection of photons, which were generated by the quantum nodes, at a central station located between the two nodes. These photonic modes, each entangled with the node that is storing the other portion of the entangled state, are mixed at a beam splitter (BS). If the modes are indistinguishable, the BS erases the information about their origin, i.e. the detection of photons after the BS projects the two quantum nodes onto an entangled state. The heralding of the entanglement then requires photonic modes traveling from the quantum nodes to the central station and the result of the photon detection traveling back to the nodes, i.e., a two-way communication. If the two quantum memories can store only one single mode and are connected by a communication channel of length $L_0$ and refractive index $n$, an entanglement creation trial duration is bounded by the communication time $\tau_{comm}=n \, L_0/c$, such that the overall repetition rate of the entanglement generation in one link is limited to $R=1/\tau_{comm}$ \cite{Simon2007}. For long distances, $R$ could decrease significantly. For example, for two single-mode quantum memories separated by 100 km of optical fibers the repetition rate of the entanglement generation trials is limited to $R=1/\tau_{comm}$ = 2 kHz, therefore seriously constraining the achievable entanglement rate.

The entanglement heralding rate can be significantly increased by the use of so-called multimode memories, which allow for multiplexing of the entanglement generation. A multimode quantum memory allows for the storage of various photonic modes in different degrees of freedom, e.g., temporal, frequency or spatial modes. By using a multimode memory that is able to store $N$ modes it is possible to perform $N$ entanglement creation trials during a communication time, therefore increasing the entanglement generation rate by a factor $N$, to first order \cite{Simon2007}. Moreover, the increase of entanglement rate has the beneficial side effect of relaxing the requirement on the storage time of the quantum memories \cite{Collins2007}. 

Quantum memories based on ensembles of atoms are well suited for developing multimode quantum memories. Cold atomic clouds have been so far mostly used to investigate spatially multimode memories \cite{Nicolas2014,Pu2017,Parniak2017,Lipka2021}, but time multiplexing has been demonstrated as well \cite{Heller2020}. Rare-earth doped crystals are particularly promising for the realization of massively multiplexed quantum memories. Their static inhomogeneous broadening can be used as a resource for time and frequency multiplexing, a precious capability that could be combined with spatial multiplexing. Among the many protocols that have been proposed to store photonic qubits in rare-earth doped crystals, the atomic frequency comb scheme \cite{Afzelius2009a} is naturally suited for temporal multiplexing \cite{Nunn2008,Afzelius2009a}. Temporal multiplexing is particularly attractive because it can be used in a simple manner in quantum repeater architecture, just by detecting the arrival time of the photon using a single detector.

Following the first demonstration of an AFC memory \cite{Riedmatten2008}, several proof-of-principle demonstrations have been performed with photonic qubits and single photons enabling light-matter \cite{Clausen2011,Saglamyurek2011,Rakonjac2021} and matter-matter entanglement \cite{Usmani2012, GrimauPuigibert2020, LagoRivera2021, Liu2021}. AFC spin-wave memories with on-demand read-out have also been demonstrated for photonic qubits \cite{Gundogan2015, Laplane2016a} and single photons \cite{Seri2017, Rakonjac2021}. Temporal multimodality has been shown in several experiments \cite{Usmani2010,Jobez2016,Laplane2017,Kutluer2017}. Spectral multiplexing has also been achieved by creating several AFCs at different frequencies within the inhomogeneous broadening \cite{Sinclair2014, Seri2019}.

In this paper, we analyze in detail the multimode capacity of AFC quantum memories in rare-earth (RE) doped crystals, taking into account realistic parameters. We develop a model to infer the maximum number of temporal modes for a given efficiency as a function of the coherence of the optical transition, the bandwidth of the AFC and the Rabi frequency of the control pulses in case of spin-wave storage. We also estimate the maximal number of spectral and spatial modes that could be stored using realistic parameters. In addition, we provide experimental demonstrations of multi-mode storage in \Eurp{} and \Prs{}, where we report the largest number of temporal modes stored both in the optical and spin transition to date. 

\section{Atomic frequency comb quantum memories}
\label{sec_AFC_intro_defs}

\begin{figure} [h]
	\centering
	\includegraphics[width=1\textwidth]{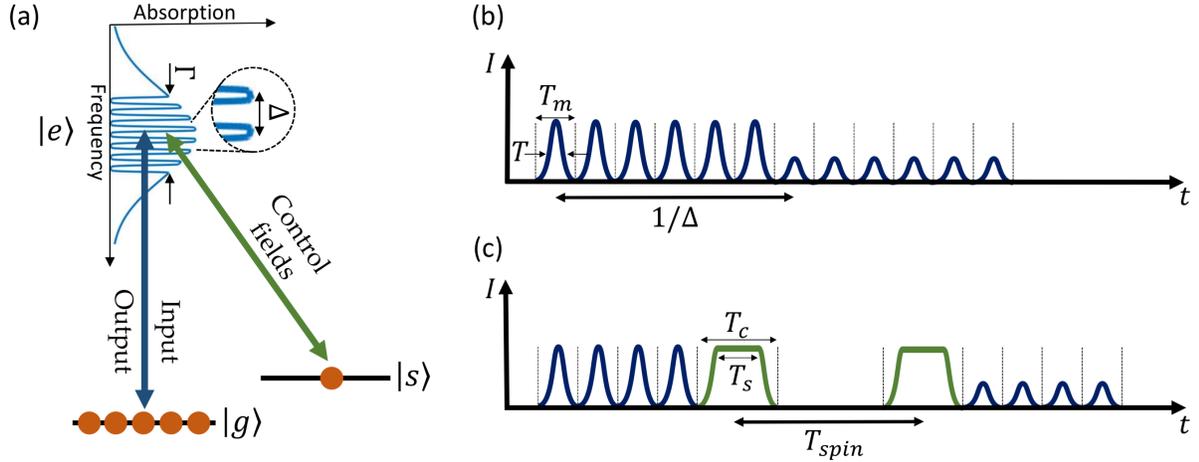}
	\caption{(a) The AFC fixed-delay memory is based on the absorption and re-emission of an echo by a comb structure of periodicity $\Delta$ and bandwidth $\Gamma$ written into an inhomogeneous absorption profile. The AFC spin-wave memory is based on a coherent and reversible transfer to a second ground state, allowing on-demand read out and long storage times. (b) Time sequence of an AFC echo memory with a fixed delay of $1/\Delta$. The input consists of a train of modes with mode duration $T_m$, each containing pulses with a FWHM of duration $T$ in intensity. Note that the train can be a sequence of pulses with random intensities, but with otherwise identical properties. (c) In the case of an AFC spin-wave memory the control pulse occupies a bin size of $T_c$, such that the total duration in which input modes can be defined is now reduced to $1/\Delta - T_c$. In this illustration the control pulse consumes exactly two input mode bins. In this article smooth adiabatic pulses with a constant intensity duration of $T_s$ are treated analytically, see text for details. The total spin-wave storage time is $T_{spin}$.}
	\label{fig_AFC_memory_scheme}
\end{figure}

Atomic frequency comb (AFC) quantum memories \cite{Afzelius2009a} are based on a periodic atomic absorption profile in the form of a comb structure, with a given periodicity $\Delta$ and total bandwidth $\Gamma$, see Figure \ref{fig_AFC_memory_scheme}(a). The comb can be created on an inhomogeneously broadened optical transition $|g\rangle$-$|e\rangle$ through the use of spectral hole burning techniques, i.e. by frequency-selective optical pumping. RE-doped crystals are ideal systems in this respect, given their large, static inhomogeneous broadening, narrow homogeneous broadening and low spectral diffusion \cite{Thiel2011}. To efficiently burn the periodic structure while maintaining high spectral resolution one can use optimized optical pumping sequences based on multi-frequency adiabatic pulses \cite{Jobez2016}.

Input pulses absorbed by the comb on the $|g\rangle$-$|e\rangle$ transition results in AFC echoes after a fixed-delay storage time of $1/\Delta$, owing to the transient response of the medium, see Figure~\ref{fig_AFC_memory_scheme}(b). If the minimum input pulse duration set by the AFC bandwidth is much shorter than the fixed storage time $1/\Delta$, then temporal multimode storage is possible. Let's consider that a smooth temporal mode occupies a truncated (cut-off) duration of $T_m$, then the temporal multimode capacity is given by $N_t = 1/(\Delta T_m)$\footnote{Formulas given throughout the article for calculating mode numbers $N_t$ will generate non-integer numbers in general. Mathematically the maximum mode number should be the integer part $\left \lfloor{N_t}\right \rfloor $, while in practice if $N_t$ is close to the next-highest integer another mode can be stored with negligible loss of efficiency.}. 
In this article we show that the multimode capacity is related to the mode size $T_m$ and the bandwidth $\Gamma$, while the exact shape and full-width at half-maximum (FWHM) of the mode is of less importance provided its total energy is mostly contained within the cut-off duration $T_m$. A detailed analysis assuming Gaussian intensity mode profiles will be given. We will also consider the impact of the optical coherence time $T_2$ (between $|g\rangle$-$|e\rangle$) on the temporal multimode capacity.

On-demand read-out and longer storage times can be achieved through the reversible transfer of the optical excitation to a second ground state $|s\rangle$ using two control fields, see Figure~\ref{fig_AFC_memory_scheme}(c). The storage time $T_{spin}$ in the spin state is limited by dephasing due to inhomogeneous spin broadening and spectral diffusion. By applying a spin echo sequence the storage time can be extended to the regime of seconds, as demonstrated when storing strong laser pulses \cite{Longdell2005,Heinze2013,Holzaepfel2020}. In the quantum regime one must deal with technical noise generated by the imperfect spin echo sequences, where up to 100~ms of spin-storage time has been achieved so far \cite{Ortu2021a}.

Efficient mapping to and from the $|s\rangle$ state requires high transfer probability over the entire bandwidth $\Gamma$ of the AFC. This is a challenge with RE ions due to their low optical oscillator strengths \cite{Thiel2011}. To circumvent this one can use adiabatic, frequency-chirped pulses \cite{Silver1985,Rippe2005,Minar2010} having a flat transfer profile over $\Gamma$. However, these pulses are much longer than the $\pi$ pulse duration set by the optical Rabi frequency $\Omega$. If the cut-off duration assigned to a single control pulse is $T_c$ (assuming two identical control pulses), then as a consequence the temporal multimode capacity is reduced to $N_t^{sw}=(1/\Delta-T_c)/T_m$. In this article we will calculate the required duration $T_c$ assuming a specific adiabatic pulse proposed by Tian et al. \cite{Tian2011a}, which is particularly efficient given a constraint of the total cut-off duration $T_c$.

\section{Theoretical temporal multimode capacity}
\label{sec_theory_temporal_MM}

\subsection{Capacity limit of the AFC fixed-delay memory due to Nyquist-Shannon sampling theorem}
\label{secsub_theory_Nyquist_AFC_delay}

The preservation of temporal information is closely related to the Nyquist-Shannon sampling theorem. As argued by Shannon \cite{Shannon1949}, a temporal signal $f(t)$ that contains a maximum frequency $W$ has its Fourier spectrum within the frequency range $-W$ to $W$, leading to the minimum required ``sampling rate'' $1/2W$. In our case information is encoded into a train of input modes separated by $T_m$ in time, meaning the maximum frequency of the time signal is $W=1/T_m$ and it requires a minimum frequency range of $2W=2/T_m$ to be accurately captured. 

\begin{figure} [h]
	\centering
	\includegraphics[width=0.67\textwidth]{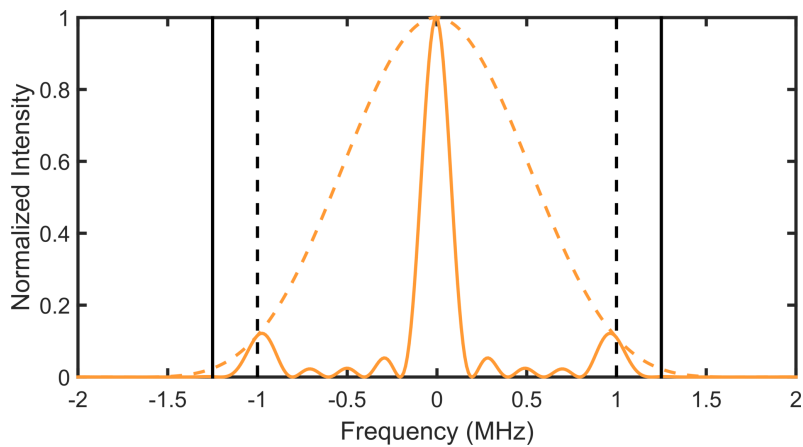}
	\caption{Fourier transform of a train of 5 Gaussian pulses of equal intensity with a FWHM of 410 ns (solid curve), each truncated to a mode size of $T_m$ = 1 $\mu$s. The Nyquist sampling limits at $\pm 1/T_m = \pm 1$ MHz are indicated by dashed vertical lines. The proposed bandwidth limit corresponding to $2.5/T_m$ is shown by the solid vertical lines. The Fourier transform of a single Gaussian mode is shown as a dashed curve.}
	\label{fig_Nyquist_gamma}
\end{figure}

This point can be illustrated by looking at the Fourier transform of a train of pulses of equal amplitude. In Fig.~\ref{fig_Nyquist_gamma} the Fourier spectrum of a sequence of 5 modes with $T_m = 1$~$\mu$s is shown. The mode spacing of $T_m$ results in the two Fourier peaks at $\pm W = \pm 1$ MHz. To preserve the information encoded at the modulation frequency $W$, then clearly one must have a total bandwidth of at least $\Gamma = 2W$, which encapsulates the Nyquist-Shannon sampling theorem. Note that a sequence of modes with random amplitude and phase modulation have its Fourier information encoded within the Nyquist limits. Hence, there is a strict minimum mode size $T_m$ imposed by the AFC bandwidth $\Gamma$, independently of the exact temporal shape of the mode inside the interval $T_m$.

Based on this argument we propose the following relation between the AFC memory bandwidth $\Gamma$ (in Hz) and the input mode size $T_m$,
\begin{equation}
	T_m = \frac{2.5}{\Gamma},
	\label{eq_T_m_limit}
\end{equation}
where we have rather arbitrarily chosen a factor of 2.5 instead of 2 to fully capture the Fourier peaks at $\pm 1/T_m$ whose widths depend on the total duration of the train of pulses. Following this definition, the temporal multimode capacity $N_t$ for a AFC fixed-delay memory is
\begin{equation} 
	N_t = \frac{1/\Delta}{T_m} = \frac{\Gamma}{2.5 \Delta}.
	\label{eq_N_t_AFC}
\end{equation}
One can also rewrite the formula as $N_t = N_{tooth}/2.5$, where $N_{tooth}$ is the number of teeth in the AFC, showing that each temporal mode requires about 2.5 additional peaks in the AFC.

The temporal mode capacity doesn't depend directly on the choice of pulse shape, nor on the FWHM of the pulse within the cut-off duration $T_m$. However, clearly one should also consider the pulse energy contained in the mode, given the mode profile. In general one can optimize the relation between the mode FWHM, $T$, and the mode size $T_m$, where we define $T_m = \kappa T$. In \ref{appendix_Gaussian_modes} the case of a Gaussian mode is treated in detail. It is shown that for $\kappa \approx 2.38$ then 99.5\% of the energy is contained both in the time cut-off $T_m$ and in the power spectrum cut-off $\Gamma$. This choice is illustrated as the dashed curve in Fig.~\ref{fig_Nyquist_gamma}. Other choices of $\kappa$ might increase the energy content in either time or frequency domain, at the expense of less energy content in the reciprocal domain. In practice any choice in the range of $\kappa = 2$ to $\kappa = 2\sqrt{2}$ preserves at least 98.1\% of the energy in either domain for Gaussian modes.

\subsection{Effects of finite optical coherence time}
\label{secsub_theory_AFC_T2}

The AFC echo memory efficiency is ultimately limited by the dephasing caused by decoherence on the optical transition. The optical coherence time, or equivalently the homogeneous linewidth, limits the AFC memory efficiency both in the optical pumping step (the AFC creation step), and during the actual storage time $1/\Delta$. State-of-the-art measurements of the AFC efficiency as a function of the fixed-delay $1/\Delta$ storage time shows exponential decays \cite{Jobez2016,Askarani2021,LagoRivera2021}. In Ref.~\cite{Jobez2016} it was argued, based on Maxwell-Bloch simulations of the AFC preparation step, that ultimately the AFC efficiency is limited by
\begin{equation}
	\eta_{T2} = \exp(-4/(\Delta \, T_2),
	\label{eq_T_2_limit}
\end{equation}
where $T_2$ is the optical coherence time of the transition \cite{Jobez2016}. Note that the relative $\eta_{T2}$ efficiency only accounts for the loss due to the optical $T_2$ limitation, given the choice of optical storage time $1/\Delta$. See Ref.~\cite{Jobez2016} for a more comprehensive discussion of the total AFC memory efficiency.

\begin{figure} [h]
	\centering
	\includegraphics[width=0.5\textwidth]{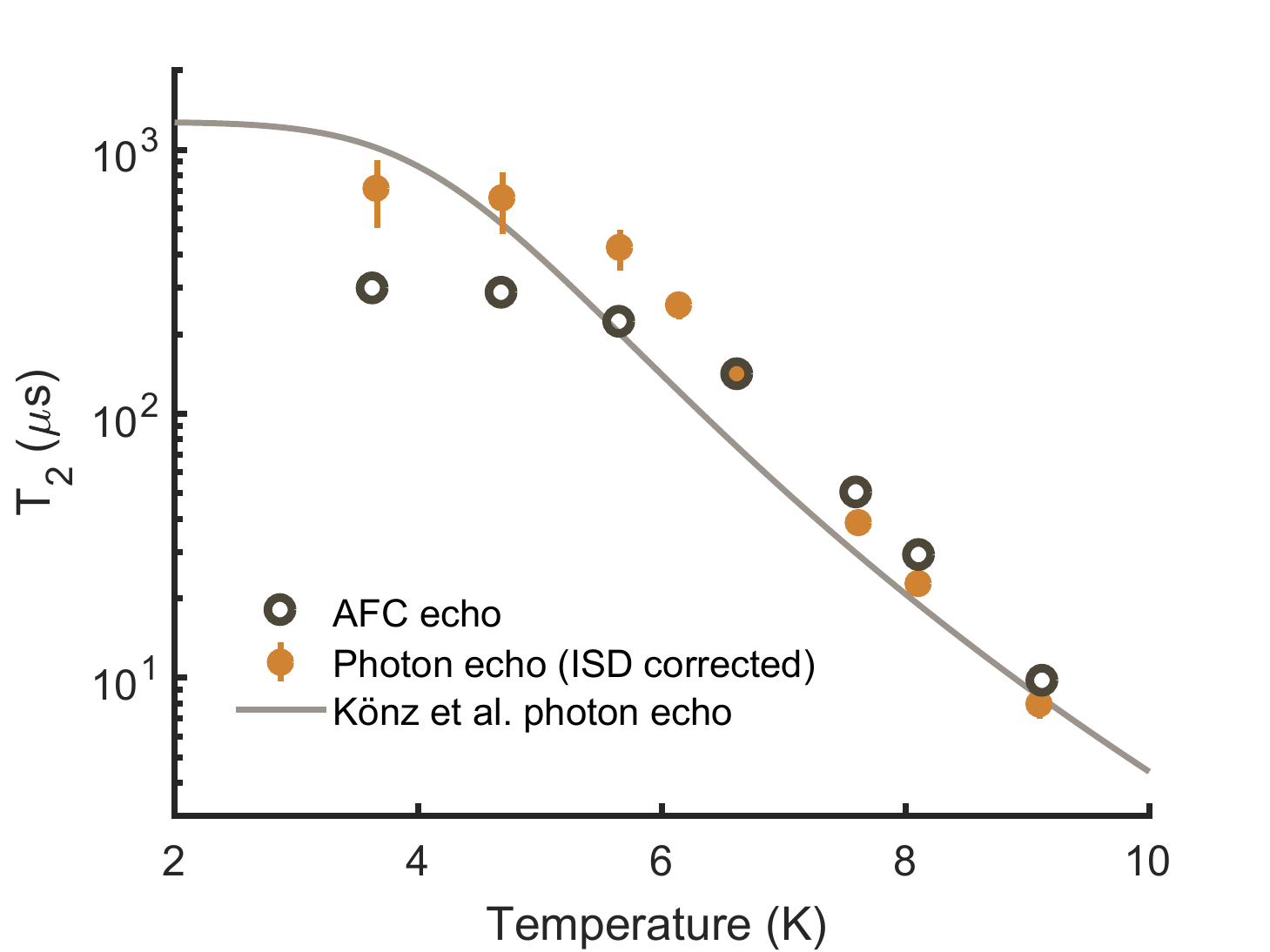}
	\caption{Coherence times obtained from AFC (open symbols) and PE (solid symbols) data in $^{151}$Eu:\yso{}, as a function of the sample temperature. The PE data has been corrected for instantaneous spectral diffusion (ISD) in the conventional manner \cite{Koenz2003}. The PE data by K\"{o}nz et al. \cite{Koenz2003} is shown for reference (solid line). The coherence times are also reported in \ref{appendix_AFC_PE_T2_EuYSO}.}
	\label{fig_PE_AFC_T2}	
\end{figure}

State-of-the-art AFC experiments \cite{Jobez2016,Askarani2021,LagoRivera2021} so far have not reached the $T_2$ limit given by Eq.~(\ref{eq_T_2_limit}). To further investigate the effect of $T_2$ we have performed both AFC efficiency decay measurements and photon echo (PE) measurements in $^{151}$Eu:\yso{} as a function of temperature, see \ref{appendix_AFC_PE_T2_EuYSO} for experimental details. The $T_2$ data from the PE measurements are taken as the reference, which the effective $T_2$ extracted from the AFC measurements should ideally reach. As shown in Figure~\ref{fig_PE_AFC_T2}, the PE and AFC coherence times do indeed converge at temperatures above 6.5 K, supporting the $\eta_{T2}$ limit introduced in Ref.~\cite{Jobez2016}. The AFC coherence time of $T_2 = 300\pm30$ $\mu$s obtained at low temperatures is the longest reported AFC coherence time so far. Yet, the PE data results in $T_2 = 707\pm204$ $\mu$s, which is a significant difference that negatively affects the current temporal multimode capacity in $^{151}$Eu:\yso{}. We believe that the lower AFC $T_2$ is due to technical issues, such as laser frequency drifts and/or dephasing induced by vibrations in the employed closed-cycle cryostats \cite{LouchetChauvet2019}. In Figure~\ref{fig_PE_AFC_T2} the temperature dependence of the $T_2$ measured by K\"{o}nz et al. \cite{Koenz2003} in a Eu-doped \yso{} sample with natural isotopic abundance is shown as reference, where a slightly longer $T_2$ of 1.27 ms was reached, which we attribute to sample differences. It should also be noted that a coherence time up to 2.6~ms has been measured with PE in a Eu-doped \yso{} sample under a weak magnetic field \cite{Equall1994}. Pr-doped \yso{} crystals generally have shorter optical coherence times, reaching 111~$\mu$s at 1.4~K at zero magnetic field and 152~$\mu$s with a weak magnetic field \cite{Equall1995}. The longest measured AFC $T_2$ of $92\pm6$~$\mu$s is close to the PE coherence time (see Sec. \ref{secsub_exp_PrYSO_AFC}), lending further support to Eq.~(\ref{eq_T_2_limit}).
 
In the following we assume that the loss of efficiency due to the optical coherence time can be modeled with Eq.~(\ref{eq_T_2_limit}), and we express $\Delta$ as a function of $\eta_{T2}$ and insert the expression into Eq.~(\ref{eq_N_t_AFC}), which gives a temporal multimode capacity of
\begin{equation} 
	N_t = \frac{\ln(1/\eta_{T2})}{10} \, \Gamma \, T_2.
	\label{eq_N_t_T2_AFC}
\end{equation}
The factor $\Gamma \, T_2$ is the ultimate upper limit of the mode capacity, as it represents the intrinsic time-bandwidth product of the memory. However, in practice only a small fraction of the memory can be exploited if an efficient storage is to be achieved. For $\eta_{T2} = 0.9$, only about 1\% of $\Gamma T_2$ can be used, while for $\eta_{T2} = 0.5$ that factor goes up to 7\% of $\Gamma T_2$ but at the cost of significantly reduced efficiency. Furthermore, it should be pointed out that $\Gamma$ is often limited by hyperfine and/or Zeeman splittings, which generally is much narrower than the entire optical inhomogeneous broadening.

\begin{figure} [h]
	\centering
	\includegraphics[width=1\textwidth]{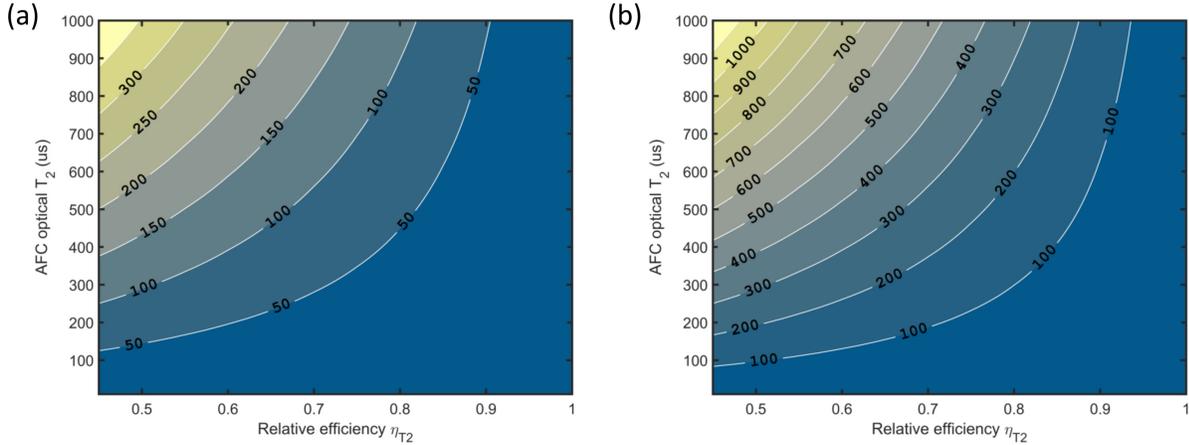}

	\caption{Contour plots of the temporal multimode capacity $N_t$ of AFC fixed-delay memories as a function of the relative efficiency $\eta_{T2}$ and the AFC optical coherence time $T_2$, calculated using Eq.~(\ref{eq_N_t_T2_AFC}) for (a) $\Gamma = 5$ MHz and (b) $\Gamma  = 15$ MHz.}
	\label{fig_map_AFC_echo_etaT2_vs_T2}	
\end{figure}

It is illustrating to plot a contour map of the AFC echo multimode capacity as a function of optical $T_2$ and relative efficiency $\eta_{T2}$, as shown in Fig.~\ref{fig_map_AFC_echo_etaT2_vs_T2}(a) and (b) for a bandwidth of $\Gamma$ = 5 MHz and 15 MHz, respectively. These bandwidths are compatible with those achievable in the RE systems $^{151}$Eu:\yso{} and Pr:\yso{} (5 MHz), and $^{153}$Eu:\yso{} (15 MHz), respectively. The plots clearly show that achieving both high multimode capacity and high relative efficiency requires long optical coherence times, given the limitations in bandwidth. 

\subsection{Multimode capacity of the AFC spin-wave memory}
\label{secsub_theory_AFC_sw}

AFC spin-wave memories require efficient transfer of the optical excitation over the entire bandwidth $\Gamma$ of the AFC. This can be achieved by chirped adiabatic control pulses \cite{Minar2010}. In NMR research, inversion pulses based on complex hyperbolic secant pulses, or sech pulses, were proposed for selective inversion of a flat frequency spectrum \cite{Silver1985}. In the adiabatic regime the sech-pulse bandwidth is entirely determined by its frequency chirp range, which follows a smooth $\tanh$ function, with a flat transfer efficiency over that bandwidth that can approach 100\% with the appropriate pulse parameters \cite{Rippe2005}. However, the sech pulse has a smooth, almost Gaussian intensity profile, hence it doesn't make very efficient use of the cut-off duration $T_c$ allocated to the control pulse. More recently Tian et al. \cite{Tian2011a} proposed an ``extended'' sech pulse, called a hyperbolic-square-hyperbolic (HSH) pulse, which has a flat intensity profile of duration $T_s$ in the center and smoothed sech pulse edges, see Figure~\ref{fig_AFC_memory_scheme}(c). The frequency chirp is still described by a $\tanh$ function, but with an extended linear regime in the center. As a result the HSH reaches significantly higher efficiency over the same bandwidth, given a cut-off duration $T_c$. The analysis here will be based on HSH control pulses, as efficient transfer using the shortest possible duration $T_c$ is paramount for the temporal multimode capacity of AFC spin-wave memories.

For the analysis it is assumed that the HSH chirp width matches exactly the AFC bandwidth $\Gamma$. We further assume that the square part of the HSH pulse $T_s$ is much longer than the edges of the HSH pulse, and that the chirp width $\Gamma$ is significantly larger than the Rabi frequency $\Omega$ of the pulse, the natural working regime for chirped, adiabatic pulses. Under these conditions the population transfer efficiency of the pulse can be written as (see Supplemental Material of Ref. \cite{Businger2020})
\begin{equation}
	\eta = 1-\exp\left( -\pi^2\frac{T_s \Omega^2}{\Gamma} \right).
	\label{eq_HSH_eff}
\end{equation}
Note that $\Omega$ and $\Gamma$ are defined in natural frequency (Hz) and not in angular frequency (rad/s) as in Ref.~\cite{Businger2020}. Now, by setting $\pi^2 T_s \Omega^2/\Gamma = 4$ it is assured that the transfer efficiency is at least 98\%. In practice it will be slightly more efficient, as the smooth sech edges of the HSH pulse will also contribute to the transfer efficiency. If we introduce the relation $T_c = \chi T_s$, where $\chi \gtrsim 1$, it follows that the spin-wave multimode capacity $N_t^{sw}$ can be expressed as
\begin{equation}
	N_t^{sw} = \frac{1/\Delta - T_c}{T_m} = \frac{1}{2.5} \left( \frac{\Gamma}{\Delta} - \chi \frac{4}{\pi^2}  \frac{\Gamma^2}{\Omega^2} \right),
	\label{eq_N_t_AFC_sw}
\end{equation}
where we have used Eq.~(\ref{eq_T_m_limit}). As it is assumed that $\Gamma > \Omega$, it follows that a certain number of modes will necessarily be consumed by the HSH pulse, where the exact number depends on the ratio $\Gamma^2/\Omega^2$. The AFC spin-wave multimode capacity can also be expressed as a function of $\eta_{T2}$ and $T_2$ to account for the finite optical coherence time, by simply modifying the first term in Eq.~(\ref{eq_N_t_AFC_sw}) to yield
\begin{equation}
	N_t^{sw} = \frac{1}{2.5} \left( \frac{\log(1/\eta_{T2})}{4} \Gamma T_2 - \chi \frac{4}{\pi^2}  \frac{\Gamma^2}{\Omega^2} \right)
	\label{eq_N_t_T2_AFC_sw}
\end{equation}

\section{Temporal multimode storage experiments in $^{151}$Eu:\yso{}}
\label{sec_exp_EuYSO}

\subsection{AFC fixed-delay multimode storage in \Eurp{}}
\label{secsub_exp_EuYSO_AFC}

Europium-doped Y$_2$SiO$_5$ features long optical and spin coherence times, hence it is particularly favorable for temporal multimodality and long-duration spin storage. Storage experiments in the quantum regime have so far utilized the $^{151}$Eu isotope \cite{Jobez2015,Laplane2017,Laplane2016a,Ortu2021a}. Given the long AFC $T_2$ obtained in the $^{151}$Eu:\yso{} system, it is particularly interesting to compare its experimental multimode storage capacity to the theoretical results obtained in Section \ref{sec_theory_temporal_MM}.

\begin{figure} [h]
	\centering
	\includegraphics[width=1.0\textwidth]{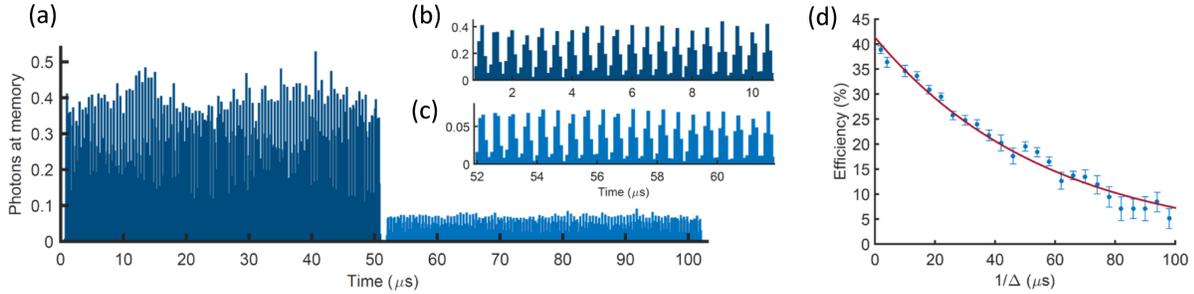}
	\caption{a) Storage of a train of 100 temporal modes in $^{151}$Eu:\yso{}, for a fixed duration of $1/\Delta = 50.7$ $\mu$s with a storage efficiency of 18\%. Vertical axis shows average number of photons at the memory. The integrated average number of photons per pulse is about 1. Panels b) and c) show a zoom on the first 20 input and output modes, respectively. d) The AFC echo efficiency as a function of storage time $1/\Delta$, resulting in an AFC coherence time of $T_2 = 249 \pm 14$~$\mu$s and a zero-time efficiency of $41.3 \pm 1.7$\%.}
	\label{fig_EuYSO_AFC_echo_exp}
\end{figure}

The bandwidth of $^{151}$Eu:\yso{} AFC memories are limited by the overlap between optical-hyperfine transitions, see for instance \cite{Lauritzen2012}, which in turn depends on the choice of three-level system used for spin-wave storage. So far AFC experiments have used either the 35 MHz or 46 MHz spin transitions, which limits the bandwidth to less than 5.7~MHz. Here we set the memory bandwidth to $\Gamma = 5$~MHz and the fixed storage time to $1/\Delta = 50.7$~$\mu$s, which results in a mode size of $T_m = 500$~ns and a temporal multimode capacity of $N_t = 100$ modes according to Eqs.~(\ref{eq_T_m_limit}) and (\ref{eq_N_t_AFC}), respectively. The intensity FWHM of the Gaussian modes were set to about 210~ns, giving a $\kappa$ parameter close to the theoretical optimum of 2.38. The mean photon number in the input modes was $\bar{n} = 0.99 \pm 0.05$, integrated over the mode size $T_m$. The experimental photon counting histograms are displayed in Figure~\ref{fig_EuYSO_AFC_echo_exp}(a-c) and the zoom on the first 20 input/output modes shows clearly distinguishable modes with these mode settings. The average storage efficiency was $18\pm2$\%.

In Figure~\ref{fig_EuYSO_AFC_echo_exp}(d) the AFC echo efficiency for a single input mode is shown as a function of $1/\Delta$. The input pulse was a bright laser pulse and the AFC echo was detected by a photodiode. The zero-time efficiency of $41.3 \pm 1.7$\% is the highest reported AFC echo efficiency without cavity enhancement. It is consistent with the measured peak optical depth of $d=5.8$, which gives an optimal theoretical efficiency of 40.1\% \cite{Bonarota2010}. The AFC $T_2 = 250$~$\mu$s is slightly shorter than the value reported in Figure~\ref{fig_PE_AFC_T2}, which we attribute to larger sample vibrations in this experiment. The relative efficiency at $1/\Delta = 50$~$\mu$s is then $\eta_{T2} = 0.45$. Given the bandwidth limitation of $\Gamma = 5$~MHz, one can store $N_t = 13$ modes at a higher relative efficiency of $\eta_{T2} = 0.9$, and $N_t = 28$ modes at $\eta_{T2} = 0.8$, according to Eq.~(\ref{eq_N_t_T2_AFC}).

\subsection{Current and future spin-wave multimode capacity in \Eurp{}}
\label{secsub_exp_EuYSO_strategy}

Multimode AFC spin-wave storage experiments have been performed under different experimental conditions in $^{151}$Eu:\yso{} \cite{Jobez2015,Jobez2016,Laplane2016a,Ortu2021a}. Multimode storage of bright coherent input modes containing large numbers of photons has reached up to 50 modes \cite{Jobez2016}. The experiment involved a storage time $1/\Delta = 41$~$\mu$s, and the AFC coherence time was relatively short at $T_2 = 110$~$\mu$s, as compared to the state-of-the-art values reported here and in \cite{Ortu2021a}. In addition, the HSH was too short, $T_c = 14$~$\mu$s, for the memory bandwidth of $\Gamma = 5$~MHz, resulting in a poor transfer efficiency of 55\% per HSH pulse. These factors contributed to an efficiency of only 1.6\%. The theoretical capacity with these values is 54 modes, according to $N_t^{sw} = (1/\Delta - T_c)/T_m$ (see Eq.~(\ref{eq_N_t_AFC_sw})) and Eq.~(\ref{eq_T_m_limit}), in close agreement with the experiment reported in \cite{Jobez2016}.

Storage of weak coherent input pulses with a mean photon number of around 1 generally requires higher efficiencies given the read-out noise of the memory, particularly when spin-echo and dynamical decoupling techniques are employed to achieve long storage times \cite{Jobez2015,Jobez2016,Ortu2021a}. Refs.~\cite{Jobez2015,Laplane2016a} showed storage of 5 temporal modes using the 35~MHz spin transition in $^{151}$Eu:\yso{}, with a spin-storage time of about 1~ms. Recently, storage of 6 temporal modes for a duration of $T_{spin} = 20$ to 100~ms was demonstrated using the 46~MHz transition, based on dynamical decoupling (DD) of the spin transition. The main difficulty in combining spin-wave storage and DD of the spin transitions lies in the read-out noise generated by errors in the DD sequence \cite{Jobez2015, ZambriniCruzeiro2016, Ortu2021a}. To achieve high signal-to-noise ratio in this context requires a spin-wave storage efficiency in the range of 5-10\%, which in turn reduces the multimode capacity. 

The theoretical spin-wave capacity can be compared to the latest experiment in Ref.~\cite{Ortu2021a}. The AFC parameters were $1/\Delta = 25$~$\mu$s and $\Gamma = 1.5$~MHz. The bandwidth was smaller than the maximum limit of 5~MHz, in order to optimize the HSH transfer pulse efficiency given the limited Rabi frequency of $\Omega = 230$~kHz, obtained with about 500 mW of power before the cryostat. The experimentally-optimized HSH pulse had parameters $T_c = 15$~$\mu$s and $T_s = 11$~$\mu$s (i.e. $\chi = 1.36$). Using these values Eq.~(\ref{eq_N_t_AFC_sw}) predicts a storage capacity of $N_t^{sw} = 5.6$ modes, in accordance with the experimentally optimized value of 6 modes. The small difference certainly lies in the strong dependence of $N_t^{sw}$ on the Rabi frequency, which is the parameter with the largest experimental error. For example, with $\Omega = 250$~kHz one finds $N_t^{sw} = 7.1$ modes. The spin-wave storage efficiency at $T_{spin} = 20$~ms was $\eta_{sw} = (7.39 \pm 0.04) \%$.

It is clear that the Rabi frequency seriously limits the currently achievable multimode capacity for AFC spin-wave memories in $^{151}$Eu:\yso{}. Currently the control pulse is applied on the weak transition of the lambda system \cite{Jobez2015,Jobez2016,Laplane2016a,Ortu2021a}, in order to favor the input transition in terms of optical depth. However, it is known that impedance-matched cavities can achieve 100\% memory efficiency using weak input transitions \cite{Moiseev2010a,Afzelius2010a,Sabooni2013,Jobez2014,Davidson2020}. Hence, a way forward is to use the strong transition for the control pulse, and use a cavity to compensate the low absorption on the input transition. According to the table of transition strengths in \Eurp{} \cite{ZambriniCruzeiro2018a}, this approach would boost the HSH Rabi frequency by a factor of 2.7, giving about $\Omega = 620$~kHz. 

\begin{figure} [h]
	\centering
	\includegraphics[width=.9\textwidth]{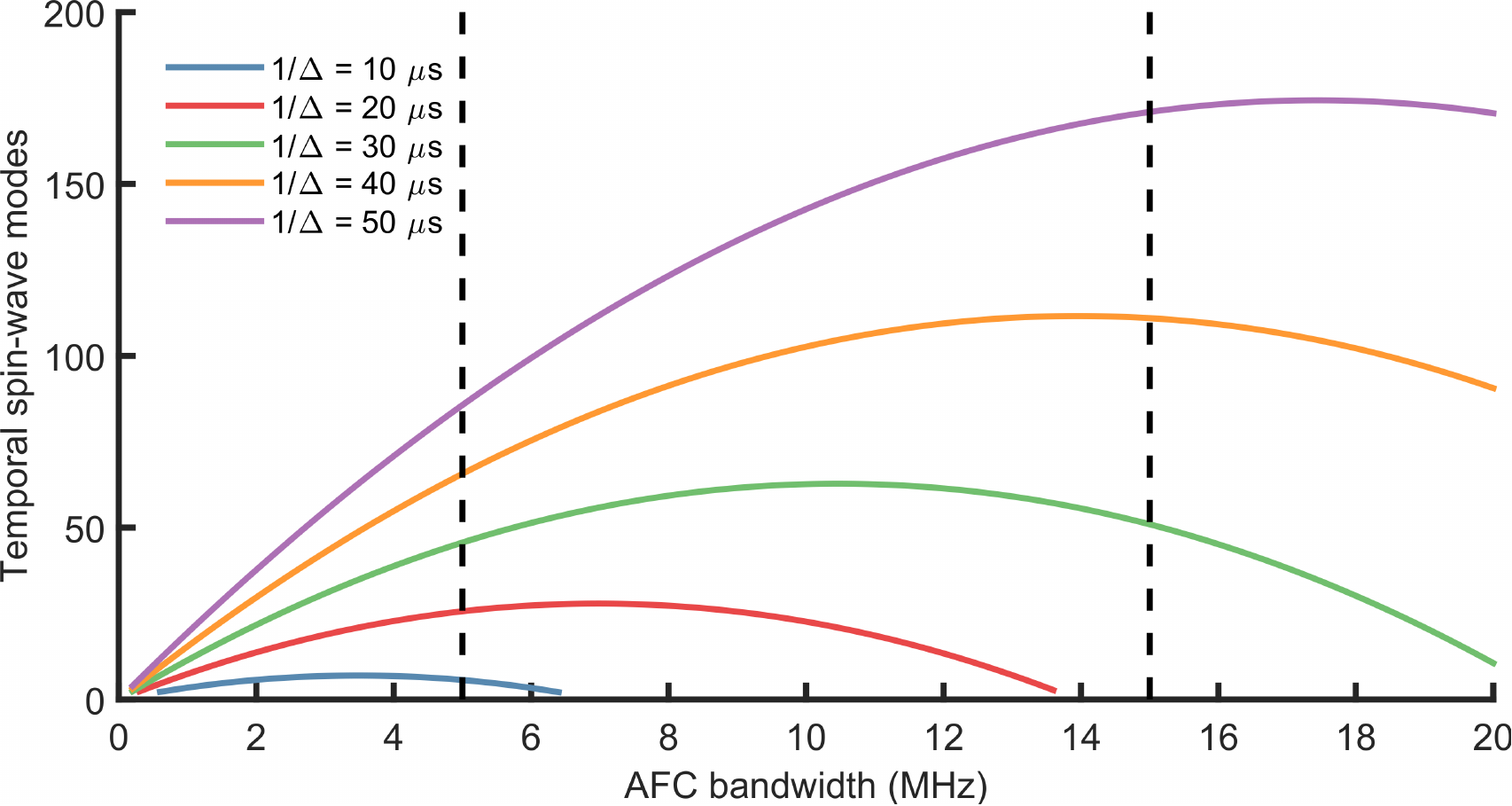}
	\caption{AFC spin-wave mode capacity $N_t^{sw}$ as a function of bandwidth $\Gamma$ according to Eq.~(\ref{eq_N_t_AFC_sw}), calculated using parameters $\Omega = 620$~kHz, $\chi = 1.36$, and $1/\Delta = 10$~$\mu$s to 50~$\mu$s by steps of 10~$\mu$s. The left and right dashed, vertical lines at 5 and 15~MHz indicate the upper bandwidth limits of $^{151}$Eu, Pr (5 MHz) and $^{153}$Eu (15 MHz) in \yso{}.}
	\label{fig_N_sw_modes_vs_Gamma}
\end{figure}

In Fig.~\ref{fig_N_sw_modes_vs_Gamma} the spin-wave multimode capacity is plotted as a function of memory bandwidth, over a range of excited state storage times $1/\Delta$, for a higher HSH Rabi frequency of $\Omega = 620$~kHz. In general there is a maximum mode capacity for an optimum bandwidth $\Gamma$, given $\Omega$ and $1/\Delta$, according to Eq.~(\ref{eq_N_t_AFC_sw}). However, one also needs to consider the maximum bandwidth supported by the physical system. The dashed lines in Fig.~\ref{fig_N_sw_modes_vs_Gamma} show the approximately maximum bandwidths achievable in $^{151}$Eu:\yso{}, \Prs{} (5~MHz) and $^{153}$Eu:\yso{} (15~MHz), respectively. It follows from this plot that with an increased Rabi frequency, for $^{151}$Eu:\yso{} and \Prs{} the multimode capacity is chiefly limited by the system bandwidth. Reaching a temporal multimode capacity of 100 or more with Europium would likely require shifting to $^{153}$Eu:\yso{} and using excited state storage times $1/\Delta$ of 40~$\mu$s or more. This in turn requires long AFC coherence times to simultaneously achieve high efficiencies, according to Eq.~(\ref{eq_T_2_limit}), which is in principle possible given the long optical coherence times in \Eurp{} crystals. 
Finally we emphasize that waveguide-based quantum memories could provide a paradigm shift for higher bandwidth memories, as the mode confinement allows Rabi frequencies significantly higher than 1~MHz \cite{Seri2018,Liu2020}. A particularly interesting system in this context is $^{171}$Yb:\yso{} \cite{Ortu2018}, where memory bandwidths of at least 100~MHz are possible due to the larger hyperfine splittings \cite{Businger2020}.

\section{Temporal multimode storage experiments in \Prs{}}
\label{sec_exp_PrYSO}

Compared to europium-doped \yso{}, praseodymium-doped \yso{} has shorter optical and spin coherence times. In the case of the optical transition, the inferred coherence time from literature is 111~$\mu$s at 1.4~K, and up to 152~$\mu$s with an applied magnetic field, in the most commonly used crystallographic site \cite{Equall1995}. However, the transition strength is larger, which can allow for shorter control pulses when performing spin wave storage. In the following two sections we will discuss the temporal multimode capacity of AFC fixed-delay and spin-wave memories in Pr:Y$_2$SiO$_5$.

\subsection{AFC fixed-delay multimode storage in \Prs{}}
\label{secsub_exp_PrYSO_AFC}

The longest reported AFC storage time in \Prs{} was $1/\Delta = 25$~$\mu$s in Ref.~\cite{LagoRivera2021}. The bandwidth of the memory was $\Gamma =4 $~MHz, limited by the hyperfine level spacing in Pr-doped \yso{}. Following Eq.~(\ref{eq_N_t_AFC}), the maximum number of modes that can be stored is $N_t = 40$ (where $T_m = 625$~ns). Ref.~\cite{LagoRivera2021} contained an analysis of the concurrence and heralding rate for an experiment entangling two \Prs{} memories versus the number of possible stored temporal modes. It was shown that the heralding rate increases with the number of modes stored, demonstrating the advantage of temporal multimode storage.

In the same experiment, the $T_2$ measured from the AFC storage efficiency was $92 \pm 6$~$\mu$s, which is very close to the $T_2$ measured with photon echoes in the same sample of \Prs{}. The resulting relative efficiency for $1/\Delta = 25$~$\mu$s is then $\eta_{T2} = 0.34$. The application of a magnetic field and lower temperatures would be needed to increase the measured $T_2$ and to reach longer storage times. Improvements in efficiency are also still possible, which are beneficial to obtaining high rates of entanglement distribution.

\subsection{Spin-wave storage in Pr:Y$_2$SiO$_5$}
\label{secsub_exp_PrYSO_AFCsw}

The advantage in using \Prs{} for temporal multimode storage becomes more apparent when performing spin-wave storage. The higher optical transition dipole moment produces a higher Rabi frequency for the same optical intensity, as compared to Eu, thereby reducing $T_c$ and thus allowing a larger temporal mode capacity given a similar $\Gamma$ and $\Delta$. 

We here present experimental results of temporal multimode spin-wave storage using \Prs{}. The experimental setup is similar to that of Ref.~\cite{Rakonjac2021}, except we used attenuated laser pulses (weak coherent states) as input modes. These had a Gaussian intensity profile, where the FWHM was $T = 180$~ns and the input mode size was $T_m = 620$~ns, resulting in $\kappa = 3.4$. The average number of photons per mode was $0.90 \pm 0.05$. The AFC delay was set to $1/\Delta = 25$~$\mu$s (the maximum storage time reported in the previous section), for which the AFC echo had an efficiency of $(7.7 \pm 0.3)\%$. The AFC bandwidth was $\Gamma = 4$~MHz, and the control pulse Rabi frequency was $\Omega = 410$~kHz with a measured power outside the cryostat of 8.5~mW. The control pulses were not HSH pulses, but less optimal pulses with a Gaussian profile and a hyperbolic secant frequency chirp spanning 4.2~MHz, where $T_c = 5$~$\mu$s and the FWHM was 2.5~$\mu$s. The resulting transfer efficiency was $(72 \pm 2)\%$. While spin-echo techniques could be applied to \Prs{} systems to extend the storage time \cite{Heinze2013}, none are applied in the following analysis.

\begin{figure}[t]
\includegraphics[width=\linewidth]{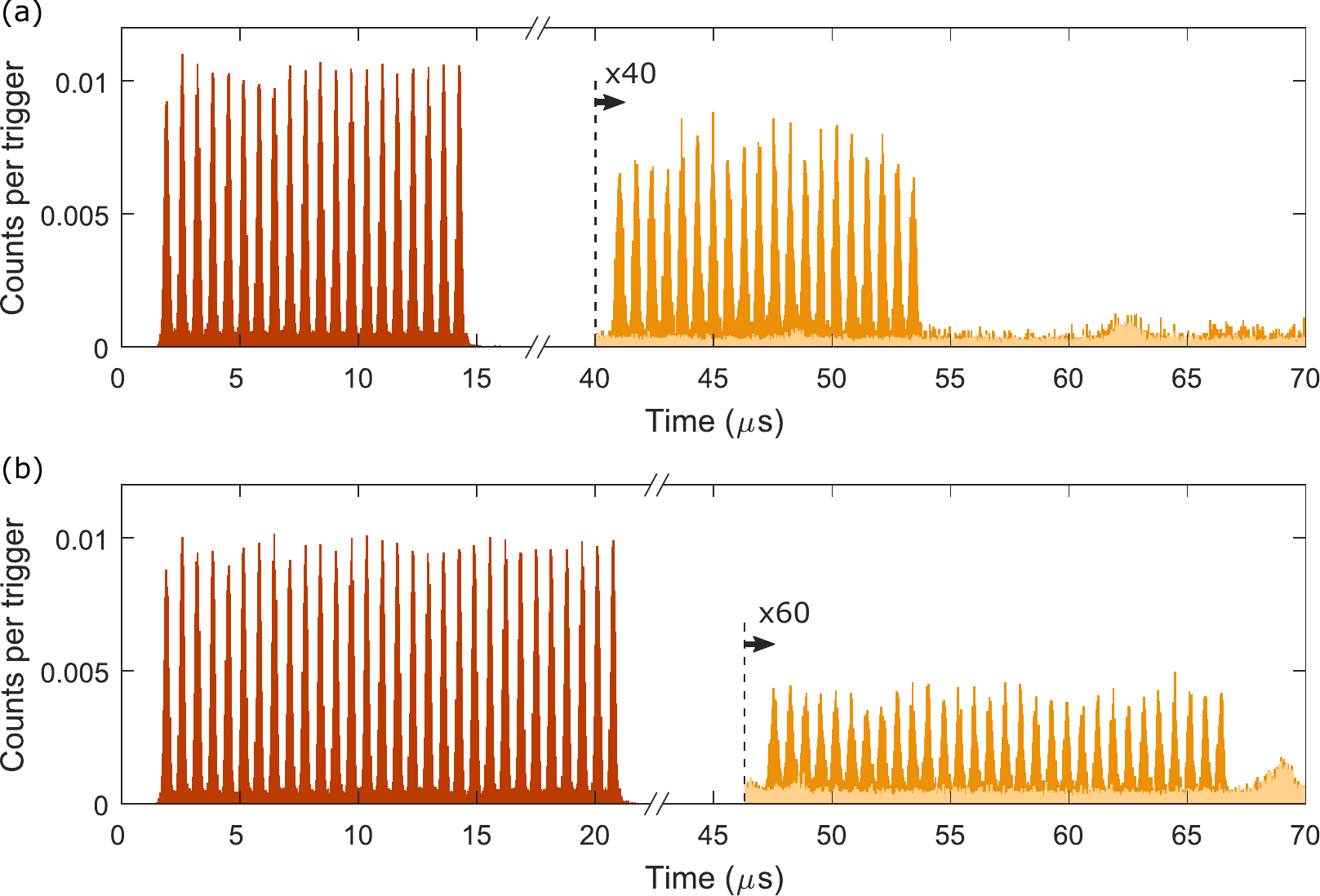}
\centering
\caption{\label{fig:mm} Input modes and retrieved spin-wave modes for (a) 20 input modes and (b) 30 input modes. The input modes measured through a transparency window in the memory are shown in brown, while the retrieved spin-wave modes are shown in orange, and the equivalent noise without storage is shown in light orange.}
\end{figure}

Figure~\ref{fig:mm} shows the retrieved temporal modes after spin-wave storage, where 20 modes were stored in (a) and 30 modes in (b). For 20 modes, the average signal to noise ratio, $\mathit{SNR}$, was $17.4 \pm 0.4$, with a corresponding spin-wave storage efficiency $\eta_{sw}$ of $(1.88 \pm 0.03)\%$. For 30 modes, $\mathit{SNR} = 6.7 \pm 0.1$, with a corresponding efficiency of $(0.63 \pm 0.01)\%$. There is a drop in $\mathit{SNR}$ and $\eta_{sw}$ when storing more modes because different $T_{spin}$ were used: 14.1~$\mu$s and 20.7~$\mu$s for 20 and 30 modes, respectively. There will be a residual signal from the AFC echoes, due to the finite efficiency of the control pulses, so $T_{spin}$ has a minimum length required to avoid the signal from the spin-wave output modes overlapping with the residual AFC echoes, which depends on the number of modes stored. The efficiency follows $\exp(- \frac{(\pi T_{spin} \gamma_{spin})^2}{2 \ln(2)})$, where $\gamma_{spin}$ is the inhomogeneous broadening of the transition used for spin-wave storage, so $\eta_{sw}$ decreases for longer $T_{spin}$. In this measurement, $\gamma_{spin}$ was measured to be 26.3~kHz. We can reduce this to 16.1~kHz in more ideal experimental conditions \cite{Rakonjac2021}, which would increase the average efficiencies to approximately 3.5\% and 2.4\% for 20 and 30 modes, respectively.

Given the $T_c$ of 5~$\mu$s, the ideal $T_m$, and the AFC storage time $1/\Delta = 25$~$\mu$s, it should be possible to store $N_t^{sw} = 32$ modes, in agreement with the 30 modes stored in Fig.~\ref{fig:mm}(b). However, the control pulses have neither the ideal profile or optimum transfer efficiency. Using Eq.~(\ref{eq_N_t_AFC_sw}), which does assume a higher transfer efficiency (and taking $\chi = 1.36$ like the Eu experiments), the maximum number of possible modes is 19. Although we can currently store more modes by using control pulses with a lower transfer efficiency, longer pulses are required to increase the total storage efficiency. Nonetheless, this still shows the benefit of using materials allowing strong Rabi frequencies for temporally multimode spin-wave storage.

\section{Multiplexing in other degrees of freedom}
\subsection{Spectral multimodality}

The wide inhomogeneous broadening of RE-doped crystals offers a significant intrinsic advantage for spectral multimodality. The inhomogeneous broadening, which allows for the implementation of the AFC protocol, arises from a static effect that increases the absorption spectrum of the ensemble, widening it by several orders of magnitude if compared to the absorption line of each single ion. For example, in \Prs{} it results in an increase in absorption from about 1 kHz to 10 GHz, while the bandwidth of the AFC quantum memory is only 4~MHz.
In \Prs{}, frequency multiplexing has been demonstrated in the case of AFC fixed-delay memories, where 15 frequency bins of a photon pair, separated by 261~MHz and spanning almost 4~GHz, have been stored simultaneously in a waveguide-integrated quantum memory \cite{Seri2019}. The storage of weak coherent states multiplexed in 26 frequency modes in a Ti:Tm:LiNbO$_3$ waveguide has also been demonstrated, followed by feed-forward-controlled frequency manipulation \cite{Sinclair2014}, as part of a proposal of quantum repeaters based on frequency multiplexing and AFC fixed-delay memories. A demonstration for spin-wave storage has also been realized, with the storage of weak coherent states in two spectral modes separated by 80~MHz \cite{Yang2018}.
\begin{figure}[t]
    \centering
    \includegraphics[width=1\textwidth]{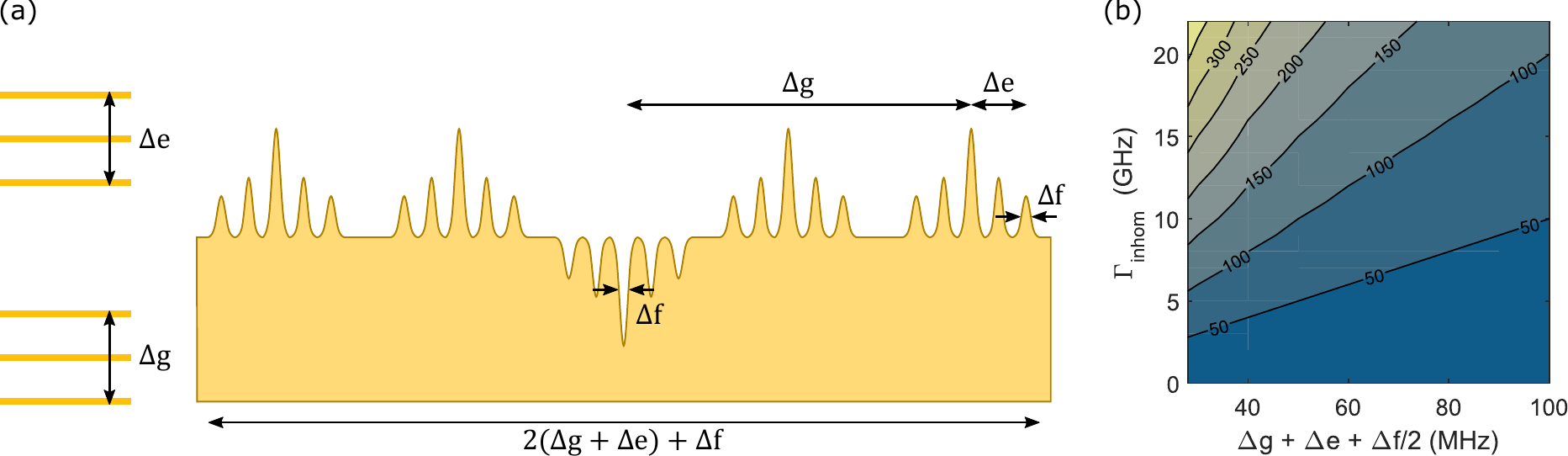}
    \caption{(a) Sketch illustrating the minimum spacing required to prepare independent features of width $\Delta f$ in the inhomogeneously broadened profile of RE ions, with total hyperfine splitting of the excited and ground state levels of $\Delta e$ and $\Delta g$, respectively. The preparation of any spectral feature, for example a spectral hole as illustrated, results in a series of holes and antiholes appearing at either side of the original one, with the same width $\Delta f$. The furthest ones will then appear at a spacing of $\Delta e + \Delta g$. The height of the side features depends on the branching ratio of the specific transitions, which are omitted in this figure. (b) Maximum number of frequency modes $N_m^f$ that can be stored in a RE-doped crystal, as a function of the inhomogeneous broadening width $\Gamma_\mathit{inhom}$ and the spacing between the different AFCs.}
    \label{fig:drawing}
\end{figure}

The maximum number of spectral modes that can be stored in the inhomogeneous profile of RE-doped crystals is limited by the number of independent AFCs that can be realized. This can be quantified by considering the specific energy level structure of the ions. The position of holes and anti-holes during spectral hole burning in the inhomogeneous profile will be dictated by the spacing between hyperfine levels, with the most distant ones appearing at frequencies $\Delta g+\Delta e$ and $-(\Delta g+\Delta e)$ from the central hole \cite{Rippe2005}, where $\Delta g$ ($\Delta e$) are the total hyperfine splitting of the ground (excited) level. If the spectral width of the feature that we are generating is $\Delta f$, then this feature will involve ions within the frequency range $\pm (\Delta g + \Delta e + \Delta f/2)$. The actual width $\Delta f$ varies depending on the system, and can be as small as the AFC width $\Gamma$ (like in Eu), or larger due to the details of the optical pumping sequence (like in Pr). Therefore, independent AFCs each dedicated to a different frequency mode will have to be realized in the inhomogeneous broadening at a distance larger than $2(\Delta g+\Delta e)+\Delta f$ between each other (see Fig.~\ref{fig:drawing}), so to avoid addressing the same ions and thus degrading the quality of one AFC while preparing the others. For a square inhomogeneous broadening of width $\Gamma_{inhom}$, the number of frequency modes that can be stored, $N_f$, shown in Fig.~\ref{fig:drawing}(b), can be defined as
\begin{equation}
	N_f  = \frac{\Gamma_\mathit{inhom}}{2(\Delta g + \Delta e) + \Delta f}.
	\label{eq_N_f}
\end{equation}
This approximation is valid if only the central portion of the inhomogeneous broadening of a RE-doped crystal is considered. The specific profile of the inhomogeneously-broadened absorption and its maximum value will determine a different maximal optical depth (OD) available for the realization of each AFC according to their position. The OD decreases with distance from the centre of the absorption band, therefore limiting the maximum efficiency achievable for the different frequency modes. The maximal efficiency reachable for different ODs follows the formula \cite{Bonarota2010}:
\begin{equation}
\label{eq:OD}
    \eta_\mathit{AFC} = (1-e^{-\tilde{d}})^2 \, \mathrm{sinc}^2(\pi/F)
\end{equation}
where $\tilde{d} = \mathit{d}/F$ and $F$ is the finesse of the comb and $d$ is the OD. Eq.~(\ref{eq:OD}) is valid for a comb with square teeth, which provides the highest efficiency, and backward retrieval from the crystal. This could be implemented by performing spin-wave storage with counter-propagating control pulses \cite{Afzelius2009a}. Note that backward retrieval can also be reached using an impedance-matched cavity \cite{Moiseev2010a,Afzelius2010a}, but the calculations for the efficiency in that case would require a more complex treatment.

\subsubsection{Discussion for \Prs}

\begin{figure}[t]
   \centering
   \includegraphics[width=1\textwidth]{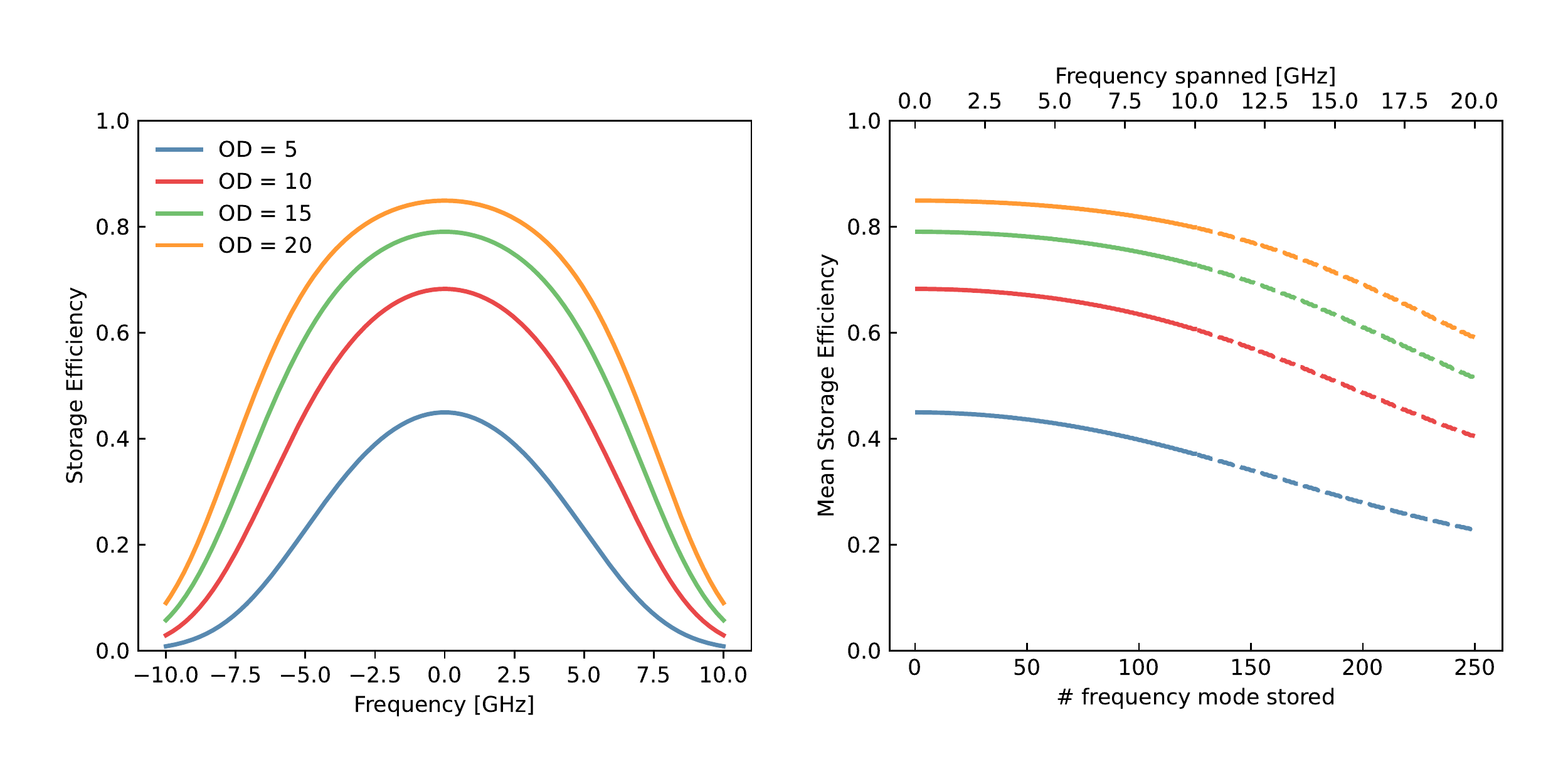}
  \caption{Efficiency variation with frequency multimodality in a \Prs{} system. (a) Change in maximum efficiency reachable at different positions in a Gaussian inhomogeneous broadening of FWHM of 10 GHz, for different maximum ODs. The efficiency curve is flattened at the centre for higher OD, allowing for a more homogenous storage in the centre of the inhomogeneous broadening. (b) Average storage efficiency as a function of the number of stored modes, for the same OD as panel (a). Higher ODs allows for a slower decay in average efficiency. The curves become dashed when the FWHM of the inhomogeneous broadening is reached.}
  \label{fig:PrYSO}
\end{figure}

The OD of a RE-doped crystal can be calculated as $d = \alpha\, L$, where $\alpha$ is the absorption coefficient and $L$ the length of the crystal. In our 5~mm long \Prs{} crystal with a doping concentration of $0.05\%$, we regularly measure an OD of 10, corresponding then to an $\alpha = 20/$cm in the center of the inhomogeneous broadening \cite{Rippe2005}. The inhomogeneous broadening in this type of crystal is well described by a Gaussian distribution with a FWHM of $\sim10$~GHz in OD \cite{Corrielli2016,Seri2018}. From the level scheme of \Prs, $\Delta g+\Delta e = 36.9$~MHz. However, this is valid only for an independent AFC, and storage in the spin-wave requires additional room for the realization of single-class features and the application of control pulses \cite{Afzelius2010}. Therefore, the width of the spectral feature to be considered is larger than the bandwidth of the AFC: for \Prs\ we have to consider $\Delta f = 18$~MHz, and therefore $2(\Delta g+\Delta e)+\Delta f \approx 92$~MHz.
The maximum efficiency achievable at different positions in the inhomogeneous broadening, calculated according to Eq.~(\ref{eq:OD}), is reported in Fig.~\ref{fig:PrYSO}(a). Panel (b) instead reports the mean storage efficiency for an increasing number of stored modes, considering also different maximal ODs of the crystal. This is naturally decreasing, as more modes with lower efficiency are stored in the wings of the inhomogeneous broadening. However, it is important to notice that while the mean efficiency is decreasing, the rate at which photons can be stored in the crystal increases linearly with the number of modes \cite{Seri2019}. Moreover, the decrease in efficiency could be counter-balanced by the use of an impedance-matched cavity, where near-unity values of efficiency can be reached even for very low values of OD and individually adjusted for every frequency mode.

\subsubsection{Discussion for \Eurp}
In the case of \Eurp{} the frequency multiplexing capability is lower than for \Prs{}. The two relevant isotopes have much larger hyperfine splittings, which spans $\Delta g+\Delta e = 258$ MHz for $^{151}$Eu \cite{ZambriniCruzeiro2018a} and $\Delta g+\Delta e = 663$ MHz for $^{153}$Eu \cite{Jobez2014} (for the commonly used site 1). Most recent quantum memory experiments in $^{151}$Eu:\yso{} have been based on isotopically pure $^{151}$Eu crystals with 1000 ppm doping concentration \cite{Ferrier2016}, where the inhomogeneous broadening is about 1.6 GHz \cite{Laplane2016a}. The frequency multimode capacity is thus at most 3 modes for $^{151}$Eu in \yso{}. However, there is a prospect of increasing the inhomogeneous broadening by introducing compositional disorder through co-doping, as observed in erbium-doped \yso{} crystals \cite{Boettger2008,Welinski2016b}. There are however open questions as to the preservation of optical and spin coherence times in this procedure.

\subsection{Spatial multiplexing}
Spatial multimodality is another precious asset available to any RE-doped crystals, that can be employed to significantly increase the number of modes available. Spatial multimodality is particularly appealing as any additional spatial mode stored would not result into a decrease in efficiency, as was the case in temporal and frequency multimodality (without an impedance-matched cavity). In the former, either the $T_2$ of the system or the separation between the modes would decrease the total efficiency, while in the latter the variation of OD across the spectral modes would result in a different maximal efficiency. In contrast, the OD is flat for parallel beam paths across the surface of the RE-doped crystal, resulting in a constant storage efficiency. This property has already been demonstrated by using two portions of the same \Prs{} crystal to store a polarization qubit \cite{Gundogan2012}. Atomic clouds behave differently, in that each spatial mode experiences a different OD, depending on its relative position to the center of the cloud. At the same time, atomic clouds do not show a static inhomogeneous broadening, meaning that they cannot take advantage of spectral multiplexing, and have only limited access to temporal storage due to the bad scaling in terms of optical depth \cite{Nunn2008}. For this reason the spatial degree of freedom has been largely exploited in such systems \cite{Lan2009, Pu2017, Chrapkiewicz2017, Tian2017, Parniak2017, Vernaz-Gris2018, Cao2020, Lipka2021}.

In RE-doped crystal, different spatial modes could be realized using for example electro-optical deflectors addressing different positions in the crystal, following a similar approach as in atomic clouds \cite{Pu2017, Chang2019, Li2021}. Further exploiting the solid-state nature of the crystal, a matrix of waveguides could be generated along the length of the sample, where different memories are addressed using commercial fiber arrays, which are at a distance of $127\, \mu$m between each other. This would result in 62 modes per mm$^2$ of crystal surface, allowing extremely high level of multiplexing. Note that this is not a fundamental limitation: a waveguide-based in- and out-coupling optical chip could collect and tightly pack modes from optical fibres with low bending loss waveguides. The chips could then be fabricated to terminate into a much denser matrix, increasing the number of modes in the crystal by more than one order of magnitude and limited only by evanescent coupling among the waveguides. 

For quantum repeater operations \cite{Simon2007}, each of the spatial modes could be addressed by an independent photon source, but this solution would make the scaling challenging. This limitation only affects the case of an external source, while would not be a problem for emissive memories \cite{Pu2017, Kutluer2017, Laplane2017}. However, even for absorptive memories the number of independent sources could be drastically reduced by using switches that could direct photons from a single source to different spatial modes at different times, translating spatial into temporal multiplexing. Another open challenge with RE-doped crystals is to combine addressing of various spatial modes with cavity-enhanced quantum memories to reach high storage efficiencies. Cavities could in principle be incorporated into waveguide arrays, but more work is needed to reduce the coupling losses to such a device.

In the discussion above we addressed the spatial multimodality obtained by realizing copies of the same quantum memories across one crystal. There are also other ways to exploit spatial multiplexing that rely on using orthogonal sets of spatial distributions of the phases of the photons along the same optical path. These include the storage of twisted light, such as optical-angular momentum states \cite{Nicolas2014, Ding2015, Zhou2015, Yang2018} or vector vortex beam \cite{Parigi2015}, or of Hermite-Gaussian or Laguerre-Gaussian modes \cite{Richardson2013}. So far, besides the usual linear polarization states, only orbital angular momentum states of light have been stored in RE-doped crystals \cite{Zhou2015, Yang2018}.

\section{Conclusions}
In this article we have considered a detailed model for quantifying the temporal mode capacity of AFC fixed-delay and spin-wave memories. These formulas can be used to compare different materials and experimental configurations, based on measurable experimental quantities. Comparisons with state-of-the-art experiments in both Eu- and Pr-doped \yso{} crystals demonstrate the validity of the theoretical models and further highlight both current limitations and future strategies for increasing the temporal multimode capacity. We finally considered the possibility of frequency and spatial multiplexing to further increase the multimode capacity, a key performance factor for future quantum repeaters.

Our analysis shows that extreme levels of multiplexing can be reached in rare-earth doped crystals, one of the few systems where temporal, spatial and spectral degrees of freedom could be exploited simultaneously thanks to their solid-state nature. Rare-earth doped crystals are therefore an evident and powerful candidate in quantum communication applications, where the entanglement distribution rate is directly proportional to the number of stored modes. Moreover, tens of temporal modes and tens of frequency modes have already been stored in various rare-earth ensembles, and the storage of several spatial modes could be easily envisioned. Temporal multimodality could then be used in synergy with spatial and frequency multiplexing, resulting in the storage of tens of thousands of modes in one, millimeter-sized solid-state crystal.

\section{Acknowledgments}
This work was financially supported by the European Union Horizon 2020 research and innovation program within the Flagship on Quantum Technologies through GA 820445 (QIA), by the Marie Sklodowska-Curie program through GA 675662 (QCALL), 713729 (ICFOStepstone 2) and 758461 (proBIST), by the Gordon and Betty Moore foundation through Grant No. GBMF7446 to H. d.-R., by the Governement of Spain (PID2019-106850RB-I00; BES-2017-082464), by CEX2019-000910-S [MCIN/ AEI/10.13039/501100011033], Fundació Cellex, Fundació Mir-Puig, and Generalitat de Catalunya through CERCA and by the Swiss FNS NCCR programme Quantum Science Technology (QSIT).

\appendix
\setcounter{section}{0}
\section{Gaussian temporal mode profile}
\label{appendix_Gaussian_modes}

The intensity profile of a Gaussian pulse with a temporal FWHM of $T$ can be written as
\begin{equation}
	I(t)=\exp(-4 \ln{2} \cdot t^2/T^2 ).
	\label{eq_Gpulse_time}
\end{equation}
Note that the field amplitude FWHM duration is $\sqrt{2} \cdot T$. The corresponding power spectrum is
\begin{equation}
	I(f)=\exp(-f^2 T^2 \pi^2 /\ln{2} )
	\label{eq_Gpulse_freq}
\end{equation}
where the spectral FWHM (in Hz) is
\begin{equation}
	\gamma = \frac{2 \ln{2}}{\pi}\frac{1}{T}.
	\label{eq_Gpulse_powerFWHM}
\end{equation}
Note that we work with power (or intensity), and not field amplitude. 

The goal here is to find an optimal relation between the pulse FWHM in time $T$, and the mode size $T_m$. One natural choice is to set $T_m = 2T$, for which 98.1\% of the pulse energy is contained in the mode. Another natural choice is to set $T_m$ as twice the FWHM in the field amplitude, i.e. $T_m = 2\sqrt{2}T$, for which 99.9\% of the energy is contained in the temporal mode bin. In practice any choice between these limits would yield low mode overlap in the time domain.

To find an optimal choice one can also calculate the ratio of the AFC bandwidth to the power spectrum FWHM of the input mode,
\begin{equation}
	\frac{\Gamma}{\gamma} = \frac{\pi}{2 \ln{2}}\Gamma  T =  \frac{2.5\pi}{2 \ln{2}} \frac{T}{T_m},
\end{equation}
where in the first step we used Eq.~(\ref{eq_Gpulse_powerFWHM}), and in the second step Eq.~(\ref{eq_T_m_limit}). Let's now set $T_m = \kappa T$ which gives us
\begin{equation}
	\frac{\Gamma}{\gamma}  =  \frac{2.5\pi}{2 \ln{2}} \frac{1}{\kappa}.
\end{equation}
The choice of $\kappa =2$, or $\kappa =2\sqrt{2}$, results in a ratio of $\Gamma/\gamma \approx 2.83$, or $2$, respectively. These in turn result in 99.9\% and 98.1\% of the power spectrum being contained in the AFC bandwidth of $\Gamma$. Comparing the two different choices it is clear that they either includes a larger portion in the time cut-off ($T_m$), or in the power spectrum cut-off $\Gamma$, respectively. Interestingly one can force the same energy content in the cut-off in both domains, by requiring
\begin{equation}
	\kappa = \frac{2.5\pi}{2 \ln{2}} \frac{1}{\kappa}
\end{equation}
which can be solved to yield an optimum choice of $\kappa$
\begin{equation}
	\kappa = \sqrt{\frac{2.5\pi}{2 \ln{2}}} \approx 2.38.
\end{equation}
With this choice, 99.5\% of the power is contained in the cut-offs of both domains.

The calculations above serve to demonstrate that the precise choice of $\kappa$ is not very critical, as soon as it is comprised between $\kappa =2$ and $\kappa =2\sqrt{2}$. It should also be noted that all formulas were derived assuming that the Fourier spectrum follows a Gaussian profile given by an \textit{ideal Gaussian pulse without truncation}. This is idealized, and in reality the cut-off in the time domain will slightly modify the Fourier spectrum. If $\kappa$ is chosen in the proposed region any correction factor is still low (i.e. less than 1\%). For the optimal choice of $\kappa = 2.38$ this effect is particularly negligible, and the truncated Fourier spectrum still contains 99.4\% of the pulse energy within the AFC bandwidth of $\Gamma$.

\section{AFC and PE coherence times recorded in $^{151}$Eu:\yso{}}
\label{appendix_AFC_PE_T2_EuYSO}

The effective coherence times extracted from the AFC fixed-delay storage experiments and the corresponding photon echo (PE) coherence times are shown in Table \ref{tab_AFC_PE_T2_EuYSO}, as a function of temperature. A brief description of the relevant experimental methods and parameters can be found below.

\begin{table}[t]
	\centering
	\begin{tabular}{c@{\quad}c@{\quad}c}
		{$T$}						& {$T_2^\text{PE}$}			& {$T_2^\text{AFC}$}	\\
		(K)						& ($\mu s$)			& ($\mu s$)			\\[2pt]
		\midrule[0.5pt]
		\midrule
		$3.7$					& $707\pm204$		& $300\pm30$		\\[2pt]
		$4.7$					& $651\pm172$		& $290\pm20$		\\[2pt]
		$5.7$					& $423\pm75$		& $222\pm13$		\\[2pt]
		$6.1$					& $256\pm29$		& $-$				\\[2pt]
		$6.6$					& $140\pm9$			& $140\pm3$			\\[2pt]
		$7.6$					& $38\pm2$		& $50.1\pm1.1$		\\[2pt]
		$8.1$					& $23\pm1$		& $29.0\pm0.4$		\\[2pt]
		$9.1$					& $8\pm1$		& $9.7\pm1.2$		\\
	\end{tabular}
	\caption{Photon echo and AFC coherence times in $^{151}$Eu:\yso{}.}
	\label{tab_AFC_PE_T2_EuYSO}
\end{table}

\subsection{Experimental setup}
	The \yso{} host crystal was doped with $^{151}$Eu at a concentration of $1000$~ppm, with dimensions $3\times2.5\times12 \text{mm}$ along the $(D_1,D_2,b)$ polarization extinction axes. Crystals from the same crystal boule were used in experiments presented in \cite{Jobez2015,Jobez2016,Laplane2016a,Laplane2017, ZambriniCruzeiro2018a, Holzaepfel2020}. A preparation beam ($700 \mu \text{m}$ beam waist) optically pumped the energy levels and prepared the AFC, while a separate input beam ($50 \mu \text{m}$ beam waist) was used for probing the tailored absorption profile and to create the input pulses to be stored. The intensity and temporal shape of the pulses were controlled by two independent acousto-optic modulators, driven by programmable arbitrary wave generators. A linear silicon detector was employed for the characterization of the memory with bright pulses, while a single photon avalanche detector was used for storage experiments with single photon-level input pulses.
	
	\subsection{Memory initialization and AFC optimization}
	The AFC experiment is initialized with a class cleaning procedure \cite{Lauritzen2012}, in order to select a unique energy level structure within the inhomogeneous broadening. In a second step the atomic population is polarized into one spin level. In a third step the AFC was prepared using the parallel AFC preparation technique described by Jobez et al.~\cite{Jobez2014}.
	
	For each AFC delay $1/\Delta$ the AFC preparation pulse parameters where optimized, these being the pulse amplitude, programmed AFC finesse and the number of pulse repetitions. Initially, a starting value of finesse is set close to the optimal theoretical finesse for square AFC peaks given the initial optical depth, see Ref. \cite{Bonarota2010}, with a low pulse amplitude to avoid power broadening. The number of pulse repetitions is then optimized to maximize the intensity of the emitted AFC echo. Then, the finesse is varied to further maximize the AFC echo intensity. The amplitude of the AFC preparation pulse is finally increased and the pulse repetition number decreased, in order to reduce the overall preparation time while keeping the same AFC echo intensity.
	
	\subsection{ISD compensation}
	The coherence time measured by photon echo is often affected by instantaneous spectral diffusion (ISD) caused by the inverting $\pi$ pulse \cite{Graf1998,Koenz2003}, which reduces the measured coherence time. Its effect can be reduced by lowering the overall excitation density $\rho_{ex}$ in the photon echo process. According to Könz et al.~\cite{Koenz2003} $\rho_{ex} = 3 \cdot 10^{12} I \tau \alpha$, where $I$ is the photon echo pulse intensity (in W/cm$^2$), $\tau$ the pulse duration (in $\mu$s) and $\alpha$ the absorption coefficient (in cm$^{-1}$). The effect of the ISD can be modelled as an effective broadening of the homogeneous linewidth $\Gamma_h=\Gamma_0+\Gamma_\text{ISD}$~\cite{Koenz2003}, where $\Gamma_\text{ISD}$ is typically proportional to the excitation density $\rho_{ex}$. The $\Gamma_\text{ISD}$ component can then be estimated by acquiring multiple photon echo decay curves for different values of the excitation density and making a linear fit to the data~\cite{Koenz2003}. We characterized the $\Gamma_\text{ISD}$ component as a function of $\rho_{ex}$ at the lowest temperature of 3.7~K, where phonon induced broadening (i.e. spin lattice relaxation) was negligible. The low-temperature $\Gamma_\text{ISD}$ component was subtracted from the homogeneous linewidth measured at all the higher temperatures, for a given excitation density, from which $\Gamma_0$ was obtained (note that $T_2^\text{PE} = 1/(\pi \Gamma_0)$). 

\pagebreak

\bibliographystyle{iopart-num}
\providecommand{\newblock}{}

\end{document}